\documentstyle[12pt,epsfig]{article}
\def\lsim{\raise0.3ex\hbox{$<$\kern-0.75em\raise-1.1ex\hbox{$\sim$}}}
\def\gsim{\raise0.3ex\hbox{$>$\kern-0.75em\raise-1.1ex\hbox{$\sim$}}}

\def\pom{{I\!\!P}}

\def\beq{\begin{equation}}
\def\eeq{\end{equation}}
\def\bea{\begin{eqnarray}}
\def\eea{\end{eqnarray}}
\def\bq{\begin{quote}}
\def\eq{\end{quote}}

\def\gappeq{\mathrel{\rlap {\raise.5ex\hbox{$>$}}
{\lower.5ex\hbox{$\sim$}}}}

\def\lappeq{\mathrel{\rlap{\raise.5ex\hbox{$<$}}
{\lower.5ex\hbox{$\sim$}}}}

\def\Toprel#1\over#2{\mathrel{\mathop{#2}\limits^{#1}}}

\newcommand{\rk}{\mbox{\boldmath $k$}}
\newcommand{\rkp}{\mbox{\boldmath $k^{\prime}$}}

\def\pom{{I\!\!P}}

\begin{document}
\pagestyle{empty}
\begin{center}
{\bf ON A TWO POMERON DESCRIPTION OF THE  $F_2$ STRUCTURE FUNCTION}
\\
\vspace*{1cm}
 A.I. Lengyel $^{1}$, M.V.T. Machado  $^{2,\,3}$\\
\vspace{0.3cm}
{$^{1}$ \rm Institute of Electron Physics, National
Academy of Sciences of Ukraine \\  Universitetska 21,
UA-88016 Uzhgorod, Ukraine  \\
$^{2}$ Instituto de F\'{\i}sica e Matem\'atica,  Universidade
Federal de Pelotas\\
Caixa Postal 354, CEP 96010-090, Pelotas, RS, Brazil\\
$^{3}$ \rm High Energy Physics Phenomenology Group, GFPAE,  IF-UFRGS \\
Caixa Postal 15051, CEP 91501-970, Porto Alegre, RS, Brazil}\\
\vspace*{1cm}
{\bf ABSTRACT}
\end{center}
  We perform a global fit to the inclusive structure function  considering
  a QCD inspired model based on the summation of gluon
  ladders describing the $ep$ scattering. In lines of a
  two Pomeron approach, the structure function $F_2$ has
  a  hard piece given by the  model and the remaining soft
  contributions: the soft Pomeron  and  non-singlet content. We have investigated several choices for the soft Pomeron and its implication in the data description.   In particular, we  carefully estimated the relative
  role of the hard and the soft contributions  in a  large span of $x$ and $Q^2$.
\vspace*{1mm}
\noindent

\vspace*{1cm}
\noindent
\rule[.1in]{16.5cm}{.002in}

\vspace{-2cm}
\setcounter{page}{1}
\pagestyle{plain}

\vspace{1cm}

\section{Introduction}

A great challenge in understanding the hard interactions  has been posed by the  HERA small-$x$  data \cite{MKlein} in order to describe the strong growth of the inclusive structure function $F_2$ as the Bjorken scale $x$ decreases. This feature  is also  supplemented by the  scaling violations on  the hard scale given by the photon virtuality $Q^2$. The currently high accuracy reached turns out the measurements having quite small statistic uncertainties, probing a large interval of virtualities, i.e. from a real photon to thousands of GeVs. As far the  Regge approach is concerned, at high energy the $ep$ scattering process is dominated by the exchange of the Pomeron trajectory in the $t$-channel. From the hadronic phenomenology, this implies that the structure function would present a mild increasing on energy ($s\simeq Q^2/x$) since the soft Pomeron intercept ranges around $\alpha_{\pom}(0)\approx 1.08$. Such behavior is in contrast with high energy $ep$ data, where the effective intercept takes values $\lambda_{\mathrm{eff}}\simeq 1.3-1.4$. In the Regge language this situation can be solved by introducing the idea of  new poles in the complex angular momentum  plane, for instance rendered in the multipoles models \cite{dippom,DM,Cudell}, producing quite successful data description. Other proposition  is given by  the two Pomeron model \cite{DL}, introducing  an additional hard intercept and corresponding residue. However, a shortcoming from these approaches is the poor knowledge about the behavior on virtuality, in general modeled in an empirical way through the vertex functions. An interesting mixed procedure is to fix the initial conditions for the QCD parton distributions (valence and sea/gluon) in a sufficiently large initial scale $Q_0^2 \sim 1-5$ GeV$^2$  from the Regge phenomenology and then perform the QCD  evolution up to higher virtualities \cite{csernai,soyez}.

On the other hand, the high photon virtuality allows the applicability of the QCD perturbative methods. Two main approaches emerge from the QCD formalism: the DGLAP and BFKL evolution schemes.  The DGLAP formalism \cite{DGLAP} is quite successful in describing most of the measurements on structure functions at HERA and hard processes in the hadronic colliders. This feature is even intriguing, since its theoretical limitations at high energy are well known \cite{OPEbreak}. Other perturbative approach is the BFKL formalism \cite{LOBFKL}, well established at LO level but not yet completely understood at NLO accuracy. The main issue in the NLO BFKL effects  is the correct account of the sub-leading corrections in the  all orders resummation \cite{BFKLNLO} (for a pedagogical review, see for instance \cite{Salam}). The LO BFKL approach can describe HERA structure function in a limited kinematical range, i.e. at not so large $Q^2$ and small-$x$. The main assumption at LO is that the processes described through the so-called hard Pomeron are driven by  gluon ladder diagrams with infinite rungs of $s$-channel gluons at asymptotic energies. A consistent treatment considering higher order resummations is currently in progress and should be available  soon. An open issue is whether the available energies in the nowadays colliders actually are asymptotic allowing the use of the complete BFKL  series.

In this work we perform a global  data fitting to the HERA structure function $F_2$ using as a model for the hard Pomeron the  finite sum of gluon ladders \cite{Trunkbfkl}. The model  is based on the truncation of the BFKL series keeping  only the  first few orders in the strong coupling perturbative expansion, where subleading contributions can be absorbed in the adjustable  parameters. From the phenomenology on hadronic collisions \cite{Fiore}, just  three orders, $\sim [\alpha_s\ln (1/x)]^2$,  are  enough to describe current accelerator data. The hard Pomeron model should be supplemented  by a  soft piece accounting for the non-perturbative contributions to the process. The description therefore turns out similar to the two Pomeron model \cite{DL}, with the advantage of a complete knowledge of the  behavior on $x$ and $Q^2$. Concerning the energy dependence, the hard piece has the logarithmic growth in contrast with the effective powerlike behavior on the model \cite{DL}.  The original model contains a reduced number of adjustable parameters: the normalization ${\cal N}_p$ and  the non-perturbative scale $\mu^2$ from the proton impact factor and the parameter $x_0$ scaling the logarithms on energy. More details on those parameter are addressed in the next section. In  a previous study at  Ref. \cite{PLB2002}, two distinct choices  for the soft Pomeron were analised. The resulting two Pomeron model was successful in describing  data on structure function $F_2$ and its derivatives (slopes on $Q^2$ and $x$) for $x \leq 2.5 \cdot 10^{-2}$  and $0.045\leq Q^2\leq 1500$ GeV$^2$. Here, we extend the kinematical range of fitting and the models considered for the soft Pomeron.

This paper is organized as follows. In the next section, we shortly review the main expressions for the hard piece given by the summation up to the two-rung ladder contribution. In   Sec. (3), an overall fit to the recent deep inelastic data is performed based on the hard contribution referred above supplemented by the remaining  soft Pomeron and non-singlet contributions. The role played by the hard and soft contributions are investigated in details. There, we also  draw up our conclusions.

\section{The hard contribution: summing gluon ladders}

In this section,  we review the elements needed to compute  the structure function using the finite sum of gluon ladders in the $ep$ collision, with the photon-proton center of mass energy labeled by $W$. Before that, let us shortly motivate the picture of the finite sum of gluon ladders. At finite energies, the LLA and NLLA  summation implies that the amplitude is represented by a finite sum on terms, where the number of terms increases
like $\ln s$, rather than by the solution of the BFKL integral equation. The interest in to take the
firts terms in the completete series in the  truncation is related to the fact
that the energies reached by the present  accelators are not high enough to
accommodate a big number of gluons in the ladder rungs that eventually
hadronize.  In the energy range of HERA there can be only a few real gluons produced in any scattering event. On the other hand, in the LO BFKL resummation real gluons can be emitted without any cost in energy, while in reality the production of a real gluon requires an energy equivalent to one or one and half units of rapidity. This violation of energy conservation  is probably cured in the NLO BFKL resummation or by relying on consistency constraint implementing energy conservation  in numerical simulations of LO BFKL evolution. The truncation of the whole series could be similar to this consistency condition. Corrobating the truncation  hypothesis, for example the coefficient weighting the term $\sim \ln^3 s$ turn out to be compatible with zero considering even
the Tevatron data \cite{Fiore}, in contrast with the expected from the
complete resummation.

 The proton inclusive
structure function, written in terms of the cross sections for the
scattering of transverse or longitudinal polarized photons, reads
as \cite{Predazzibook}
\begin{eqnarray} F_2(x,Q^2) &=& \frac{Q^2}{4\pi^2 \alpha_{\rm{em}}}
\left[\sigma_T(x,Q^2) + \sigma_L(x,Q^2)\right] \,,\\
\sigma_{T,\,L}(x,Q^2)& = & \frac{{\cal G}}{(2\pi)^4} \, \int
\frac{d^2\rk}{\rk^2}\,\frac{d^2\rkp}{\rkp^2}\,\Phi^{\gamma^*}_{T,\,L}(\rk)\,
{\cal K}(x,\rk,\rkp)\,\Phi_p(\rkp)\,, \label{eq1}
\end{eqnarray}
where ${\cal G}$ is the color factor for the
color singlet exchange and $\rk$ and $\rkp$ are the transverse
momenta of the exchanged reggeized gluons in the $t$-channel. The
$\Phi^{\gamma^*}_{T,\,L}(\rk)$ is the virtual photon impact
factor  and $\Phi_p(\rkp)$ is the proton
impact factor. The first one is well known in perturbation theory
at leading order, while the latter is modeled
since in the proton vertex there is no hard scale to allow
pQCD calculations. The kernel ${\cal K}(x,\rk,\rkp)$ contains the dynamics of the process, for instance, the BFKL kernel.

The amplitudes can be calculated order by order: for instance the Born contribution coming from  the two gluon exchange and the one-rung ladder contribution read as,
\begin{eqnarray}
{\cal A}^{(0)} & = & \frac{2\,\alpha_s \, W^2}{\pi^2}\,\sum_f \, e^2_f\,
\int \frac{d^2\rk}{\rk^4}\,\Phi^{\gamma^*}_{T,\,L} (\rk)\,\Phi_p(\rk)\,,\nonumber\\
 {\cal A}^{(1)} & = &  \frac{6 \alpha^2_s
W^2}{8 \pi^4}\, \sum_f\, e^2_f\, \ln \left( \frac{W^2}{W^2_0}\right)\, \int \frac{d^2\rk}{\rk^4}\,
\frac{d^2\rkp}{\rkp^4}\,
 \Phi^{\gamma^*}_{T,\,L}(\rk)\,{\cal K}(\rk,\rkp) \,\Phi_p(\rkp) \,,\nonumber
\label{eq2}
\end{eqnarray}
where $\alpha_s$ is  considered fixed since we
are in the framework of the LO BFKL approach.
The perturbative kernel  ${\cal K}(\rk,\rkp)$ can be calculated order by order in the perturbative expansion \cite{Predazzibook}.  The Pomeron is
 attached to the off-shell incoming photon through the quark loop
diagrams, where the Reggeized gluons are attached to the same and to
different quarks in the loop. The virtual photon impact factor averaged over the transverse polarizations reads as \cite{LevRysk,Balitsky},
\begin{eqnarray}
\Phi^{\gamma^*}_{T,\,L}(\rk)=\frac{1}{2} \int_0^1
\frac{d\tau}{2\pi} \,\int_0^1 \frac{d \rho}{2\pi}\,
\frac{\rk^2(1-2\tau\, \tau^{\prime})(1-2\,\rho \,\rho^{\prime})}{\rk^2
\,\rho\, \rho^{\prime} + Q^2 \rho \,\tau\, \tau^{\prime}},
\end{eqnarray}
where $\rho$, $\tau$ are the Sudakov variables associated
 with  the momenta in the photon vertex and  $\tau^{\prime}\equiv (1-\tau)$
and $\rho^{\prime}\equiv (1-\rho)$.

Gauge invariance requires  the proton impact factor vanishing at $\rkp$ going to zero and is modeled  in a simple way,
\begin{eqnarray} \Phi_p(\rkp)={\cal N}_p
\,\frac{\rkp^2}{\rkp^2 + \mu^2},
\end{eqnarray}
where ${\cal N}_p$ is the unknown normalization of the proton
impact factor and $\mu^2$ is a scale which is typical of the
non-perturbative dynamics. These scales will  be considered adjustable parameters in the analysis.  Considering the electroproduction process,
summing the  first orders in perturbation theory  we can write the
expression for the inclusive structure function,
\begin{eqnarray}
F_2^{\mathrm{hard}}(x,Q^2) & = & \frac{8}{3}\,\frac{\alpha^2_s}{\pi^2} \sum_f e^2_f \, {\cal N}_p
\left[\, I^{\,(0)}(Q^2,\mu^2) + \frac{3\,\alpha_s}{\pi}\,\ln
\frac{x_0}{x}\,I^{\,(1)}(Q^2,\mu^2)\,  \right. \nonumber
\\
& &  + \left.
 \frac{1}{2}\,\left(\frac{3\,\alpha_s}{\pi}\,\ln \frac{x_0}{x}\right)^2
\,I^{\,(2)}(Q^2,\mu^2)\, \right],
\label{eq7}
 \end{eqnarray}
where the functions $I^{\,(n)}(Q^2,\mu^2)$ correspond to the $n$-rung gluon ladder contribution. The quantity  $x_0$ gives the scale normalizing  the logarithms on energy for the LLA BFKL approach, which is arbitrary and enters as an
additional parameter. The contributions are written explicitly as,
\begin{eqnarray}
I^{\,(0)} & = &  \frac{1}{2} \ln^2 \left(
\frac{Q^2}{\mu^2} \right) + \frac{7}{6}\ln \left(
\frac{Q^2}{\mu^2} \right) + \frac{77}{18} \,,\\
I^{\,(1)} & = &  \frac{1}{6} \ln^3 \left(
\frac{Q^2}{\mu^2} \right) + \frac{7}{12}\ln^2 \left(
\frac{Q^2}{\mu^2} \right) + \frac{77}{18}\ln \left(
\frac{Q^2}{\mu^2} \right) + \frac{131}{27} + 2\,\zeta (3)\,,\\
I^{\,(2)} & = & {1\over24}\ln^4 \left(\frac{Q^2}{\mu^2}\right)+
\frac{7}{36}\ln^3 \left(\frac{Q^2}{\mu^2}\right)+ \frac{77}{36}\ln^2 \left(\frac{Q^2}{\mu^2}\right) \nonumber\\
& & + \left(\frac{131}{27} +4 \zeta(3)\right) \ln \left( \frac{Q^2}{\mu^2}\right) + \frac{1396}{81} - \frac{\pi^4}{15} + \frac{14}{3}\zeta(3)\,,
\label{eq10}
\end{eqnarray}
where  $\zeta (3)\approx 1.202$. It is worth mentioning the clear behavior on $x$ and virtuality. The main result in Ref. \cite{PLB2002} is in a good agreement, in terms of a $\chi^2/\mathrm{dof}$ test,  for the
inclusive structure function in the range  $0.045 \leq Q^2 \leq 1500$ GeV$^2$, once considering the restricted kinematical constraint $x \leq 0.025$.
 The non-perturbative contribution (from the soft dynamics), initially considered as a background,  was found  to be not negligible. In the next section, we perform a global analysis, extending the range on $x$ fitted by adding the non-singlet contribution modeled through the usual Regge parameterizations and analyzing also  the multipole  models for the soft Pomeron.

\section{Fitting results and conclusions}

In order to perform the fitting procedure, for the hard piece one uses
 Eq. (\ref{eq7}) and for the soft piece we have selected some typical  models, as addressed below. First,  one takes into account the  latest version \cite{KMP} of the CKMT model \cite{CKMT}, having a economical number of parameters,

\begin{eqnarray}
F^{\mathrm{soft}}_2(x,Q^2)= A\,\left(
\frac{1}{x} \right)^{\Delta(Q^2)} \left(
\frac{Q^2}{Q^2+a}\right)^{\Delta(Q^2)+1}\,\left(1-x\right)^{n_s({Q^2})}
\,, \label{softpom}\\
{\Delta(Q^2)}= \Delta_{0} \left( 1+
\frac{Q^2 \Delta_{1}}{\Delta_{2}+Q^2}\right)\,,
\hspace{0.6cm}
{n_s}(Q^{2})=\frac{7}{2} \,
\left(1+\frac{Q^{2}}{Q^{2}+b}\right)\,,
\label{delta}
\end{eqnarray}
where $\Delta(Q^2)$ is the Pomeron intercept. The non-singlet term  takes the  following form,
\begin{eqnarray}
 F^{\mathrm{ns}}(x,Q^2) & = & A_R\, x^{1-\alpha_{R}}\,\left(\frac{Q^2}{Q^2+a_R}\right)^{\alpha_{R}}\,(1-x)^{n_{ns}({Q^2})}\, \label{softreg} \\
 n_{ns}(Q^2) & = & \frac{3}{2}\,\left(1+\frac{Q^2}{Q^2+d}\right).
 \end{eqnarray}

The CKMT model  was originally constructed to interpolate between the soft hadronic Pomeron phenomenology, where $\alpha_{pom}\simeq 1.08$ and the growth for the low $Q^2$ proton structure function in deep inelastic scattering, where $\alpha_{pom}\simeq 1.2$. This is obtained through absorbtive corrections to the bare soft Pomeron, leading to a $Q^2$-dependent intercept. Therefore, in phenomenological application considering a two-Pomeron approach, we  should be careful in  verifying the soft Pomeron intercept coming out.

 Another possibility is to select a soft piece which corresponds to the
 Pomeron with
 intercept equal to one and has the form of non-perturbative truncated
  $\ln\left(\frac{Q^{2}}{x}\right)$ series
 (soft multipole Pomeron) \cite{DM,Cudell}, given by

\begin{equation}
 F^{\rm{soft}}_{2}(x,Q^{2})=Q^{2}
\left[ A \cdot \left( \frac{a}{Q^{2}+a} \right)^{\alpha} + B\cdot
\ln \left( \frac{Q^{2}}{x} \right) \left( \frac{a}{Q^{2}+a}
\right)^{\beta} \right]\,\label{dipole}
\end{equation}
 for the dipole Pomeron and
\begin{equation}
\label{tripole}
 F^{\rm{soft}}_{2}(x,Q^{2})=Q^{2} \left[ A \cdot \left(
\frac{a}{Q^{2}+a} \right)^{\alpha} + B\cdot \ln^2 \left(
\frac{Q^{2}}{x} \right) \left( \frac{a}{Q^{2}+a}
\right)^{\beta}\right]\,\end{equation}
for the  tripole Pomeron.

The multipole  means the soft Pomeron is considered as multiple poles instead of just a single pole. This approach has been successful in hadron-hadron and lepton-hadron (mostly vector meson photoproduction) scattering, where the increasing of the cross section on energy can be produced with a unit Pomeron intercept. The multipole model has shown to be quite stable in fitting simultaneously  hadronic and lepton-hadron data.

Concerning the hard piece, Eq. (\ref{eq7}), we selected the
 overall normalization factor as a free parameter, defined as ${\cal
N}=\frac{8}{3}\,
 \frac{\alpha^2_s}{\pi^2}\, \sum e^2_f \,{\cal N}_p$, considering
 four active flavors. That contribution  was also supplemented by a factor
  $(1-x)^{n_{s}}$ accounting for the large $x$ effects. The remaining parameters are the scales $\mu^2$ and $x_0$ as well as the coupling constant $\alpha_s$ (considered at a fixed scale in the BFKL formalism). As we will see in the following, its value presents small variation in the fitting, suggesting that it can be considered fixed as $\alpha_s=0.2$, which is typical in the HERA kinematical region.

 For the fitting
  procedure we consider the data set containing all
available HERA data for the proton structure function $F_2$
\cite{H1c3}-\cite{ZEUSc4}, updated with new HERA
experiment \cite{ZEUSc5} and  adding only the latest (E665 and NMC) data
set of fixed target experiments \cite{E665,NMC}. This choice  for the selection of data sets follows the procedures used in Ref.  \cite{DM}. The whole data set contains a total number of
$1059$ experimental points, covering the complete  available
 $x$ and virtualities $0.045\leq Q^2 \leq 30000$ GeV$^2$. It should be stressed that the inclusion of the  older fixed
target experimental data (SLAC \cite{SLAC} and BCDMS \cite{BCDSM}) requires a more
 sophisticated study for the  background, for example, following the  ALLM parameterization or as in Ref. \cite{DM}. However, for that purposes the number of free parameters would be significantly larger.

\begin{table}[t]
\begin{center}
\begin{tabular}{||l|l|l|l|l|l|l|l||}
\hline
\hline
&   &   I       &  II     & III  & & IV & V \\
\hline
\hline
  & $\cal N$ & 0.0237  & 0.0238 & 0.0195 & & 0.0269 & 0.0195 \\
Hard Pomeron &  $\mu^2$  & 1.36  & 1.40 & 0.108 & & 1.42 & 1.33 \\
& $x_0$ & 0.798$\cdot 10^{-2}$ & 1.01$\cdot 10^{-2}$ & 0.398 $\cdot10^{-2}$ &
& 0.833 $\cdot 10^{-2}$ & 0.808$\cdot10^{-2}$ \\
& $\alpha_s$ &0.202 &0.217 &0.0804 & & 0.204 & 0.215\\
\hline & A & 0.327 & 0.366 &- & A &0.291 &0.220\\
& a & 0.754 & 1.11 & - & a & 2.39 & 1.19\\
Soft Pomeron & $\Delta_0$ & 0.638$\cdot10^{-2}$ & 0.07(fixed) & - & $\alpha$ & 1.87 &1.32\\
& $\Delta_1$ & 25.7 & 6.00 & - & $B$ &0.712$\cdot10^{-2}$ & 0.437$\cdot10^{-3}$ \\
& $\Delta_2$ & 6.31 & 0.02(fixed) & - & $\beta$ & 1.20 &1.17\\
& b & 33.4 & 7.35 & 2.00& &19.6 & 39.5\\
 \hline & $A_R$ & 5.94 & 6.73 & 10.3& & 6.31 & 5.88\\
 Non-singlet & $a_{R}$ & 653 & 306 & 869 & & 502 & 650\\
  & d & 1.14 & 0.796 & 2.22 & & 0.947 & 1.02\\
\hline
$\chi^2/\mathrm{d.o.f.}$ &    & 1.10 & 1.23 & 1.39 & & 1.09 &1.03 \\
\hline
\hline
\end{tabular}
\end{center}
\caption{Parameters of the fit. Procedure (I): hard plus soft terms;
 Procedure (II): similar to analysis (I), but restricting soft Pomeron
  intercept; Procedure (III): only hard term and non-singlet contribution.
  Procedure (IV): hard + soft dipole pomeron terms; Procedure (V): hard + soft tripole pomeron terms}
\end{table}

\begin{table}[t]
\begin{center}
\begin{tabular}{||l|l|l|l|l|l||}
\hline
\hline &   &   A       & &  B     & C \\ \hline \hline
  & $\cal N$ & 0.0196  & & 0.0280 & 0.0195\\
Hard Pomeron & $\mu^2$ & 1.38 & & 1.42 & 0.143\\
& $x_0$ & 0.915$\cdot10^{-2}$ & & 0.793$\cdot10^{-2}$ & 0.940$\cdot10^{-2}$\\
\hline & A & 0.317 & & 0.134 & 0.212 \\ & a & 0.757 & & 2.74 & 1.25\\
Soft Pomeron & $\Delta_0$ & 0.0515 & $\alpha$ & 2.18 & 1.39\\
& $\Delta_1$ & 3.38 & $B$ & 0.469$\cdot10^{-2}$ &  0.513$\cdot10^{-3}$\\
& $\Delta_2$ & 9.27 & $\beta$ & 1.20 & 1.19\\
 \hline & $A_R$ & 10.2 & & 10.4 & 10.2\\
 Non-singlet & $a_{R}$ & 698 & & 112 & 543 \\
\hline $\chi^2/\mathrm{d.o.f.}$ &    & 0.95 & & 1.00 & 0.92
\\ \hline
\hline
\end{tabular}
\end{center}
\caption{Parameters of the fit for restricted domain
$x\leq0.07$. Procedure (A): hard plus soft terms; Procedure (B):
hard + soft dipole pomeron terms; Procedure (C): hard + soft tripole
pomeron terms.}
\end{table}

\begin{figure}[h]
\centerline{\epsfig{file=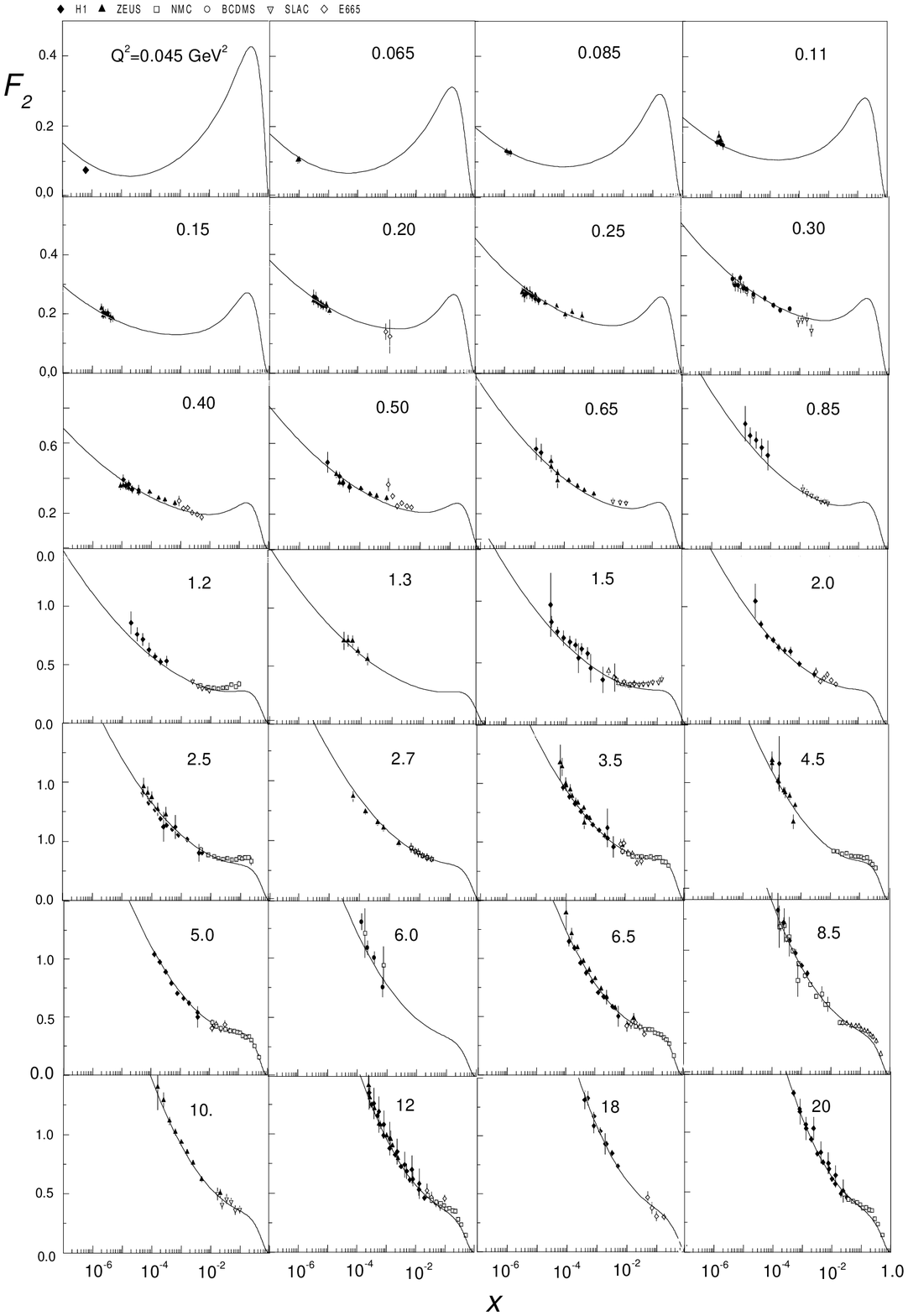,width=13cm,height=16cm}}
\caption{The inclusive structure function at small $Q^{2}$
virtualities. The procedures (I) - (V) produce nearly equivalent curves.The SLAC \cite{SLAC} and BCDMS \cite{BCDSM} data are not included into the fit. The virtualities are in units of GeV$^2$.}
\label{fig1}
\end{figure}

Here, we have considered the following distinct fitting procedures:

\begin{itemize}

\item  (I) Overall fit considering the hard piece and the CKMT soft Pomeron plus reggeon piece, Eqs. (\ref{softpom})   and (\ref{softreg});

\item  (II) Overall fit restricting the soft Pomeron intercept, modifying the expression for $\Delta (Q^2)$, see  Eq. (\ref{delta}),  in the CKMT soft Pomeron. It reads now as,
\begin{eqnarray}
 {\Delta(Q^2)}= \Delta_0 \left(
\frac{Q^2}{\Delta_1+Q^2}\right) + \Delta_2\,,
\end{eqnarray}

\item  (III) The fit is performed considering only  the hard piece plus the non-singlet
  contribution;

\item (IV and V) Overall fit using the hard Pomeron piece and the
  soft multipoles Pomeron, Eqs. (\ref{dipole}) and  (\ref{tripole}). The main feature is the soft pomeron having an intercept equal one.

\end{itemize}

    The best fit parameters for these procedures are presented in  Table (1). The kinematical range covered is
$0\leq x \leq 1$ and $0.045 \leq Q^2 \leq 30000$. In the
    following we discuss each procedure, pointing out the relative contribution
     from the hard and soft pieces and its quality.

Lets start by the procedure (I). The parameters for the hard piece remains conssistent with the previous analysis in \cite{PLB2002}, i.e. $\alpha_s\sim 0.2$, $x_0\simeq 10^{-2}$ and $\mu^2$ of order 1 GeV$^1$. Concerning the soft Pomeron, the bare intercept $\Delta_0$ coming out quite small in the whole range of $Q^2$. This fact seems corroborate a soft Pomeron having an intercept close to one in this two-Pomeron analysis. The same happens for the procedure (II), with a modified form for the CKMT intercept. For low $Q^2$ the intercept comes out small, but reaches at higher virtualities to the upper limit $\Delta =0.08$.

 The procedures (IV and V), considering the soft multipole Pomeron, provide
 the best  quality of fit  $\chi^{2}/\mathrm{dof} \simeq 1$. The tripole Pomeron is preferred as a soft background, giving a slightly better result than the dipole pomeron. Concerning the hard piece, the value for the fixed coupling constant is quite stable, $\alpha_s\simeq 0.2$, consistent with the overall HERA value.

The situation differs only in procedure (III), where a small value
for $\alpha_s$ is found,  suggesting in this case that the best
choice would be  $\alpha_{s}=0.119.$ which coincides with
$\alpha_s (M_Z)$. The high {$\chi^{2}/\mathrm{dof}$ for this case
is not quite sizeable, in view of the smaller number of parameters
considered (8 against 13 from the remaining analysis).

 Let start by the
procedure (I). The parameters for the hard piece remains
consistent with the previous analysis in \cite{PLB2002}, i.e.
$\alpha_s\sim 0.2$, $x_0\simeq 10^{-2}$ and $\mu^2$ of order 1
GeV$^2$. Concerning the soft Pomeron, the bare intercept
$\Delta_0$ comes out quite small in the whole range of $Q^2$. This
fact seems to corroborate a soft Pomeron having an intercept close
to one in this two-Pomeron analysis. The same happens for the
procedure (II), with a modified form for the CKMT intercept. For
low $Q^2$ the intercept comes out small, but reaches at higher
virtualities to the upper limit $\Delta =0.08$.

 The procedures (IV and V), considering the soft multipole Pomeron, provide
 the best  quality of fit  $\chi^{2}/\mathrm{dof} \simeq 1$. The tripole Pomeron is preferred as a soft background, giving a slightly better result than the dipole pomeron. Concerning the hard piece, the value for the fixed coupling constant is quite stable, $\alpha_s\simeq 0.2$, consistent with the overall HERA value.

The situation differs only in procedure (III), where a small value
for $\alpha_s$ is found,  suggesting in this case that the best
choice would be  $\alpha_{s}=0.119,$ which coincides with
$\alpha_s (M_Z)$. The high {$\chi^{2}/\mathrm{dof}$ for this case
is not quite sizeable, in view of the smaller number of parameters
considered (8 against 13 from the remaining analysis).

In Figs. \ref{fig1} and  \ref{fig2}, we present the fit result for the inclusive structure function for small and large virtualities. The plots for the different procedures lie on top of each other.  In Fig. \ref{fig3}, we present the  relative role of the distinct
contributions for the fit. The hard and soft pieces are presented separately.  We present them  explicitly for the
virtualities 15, 150 and 1500 GeV$^2$, where the contributions from fitting (I), (II) and (V) are shown. It is verified that the region where the hard Pomeron starts to dominate depends on the vituality. For instance, at $Q^2=15$ GeV$^2$ it stays on $x\sim 10^{-3}$ for the different procedures, whereas is shifted to $x\sim 10^{-2}$ at $Q^2=1500$ GeV$^2$.  A two-Pomeron picture is supported,  comparable with those using the two-Pomeron analysis in Refs. \cite{DM,DL}.

\begin{figure}[h]
\centerline{\epsfig{file=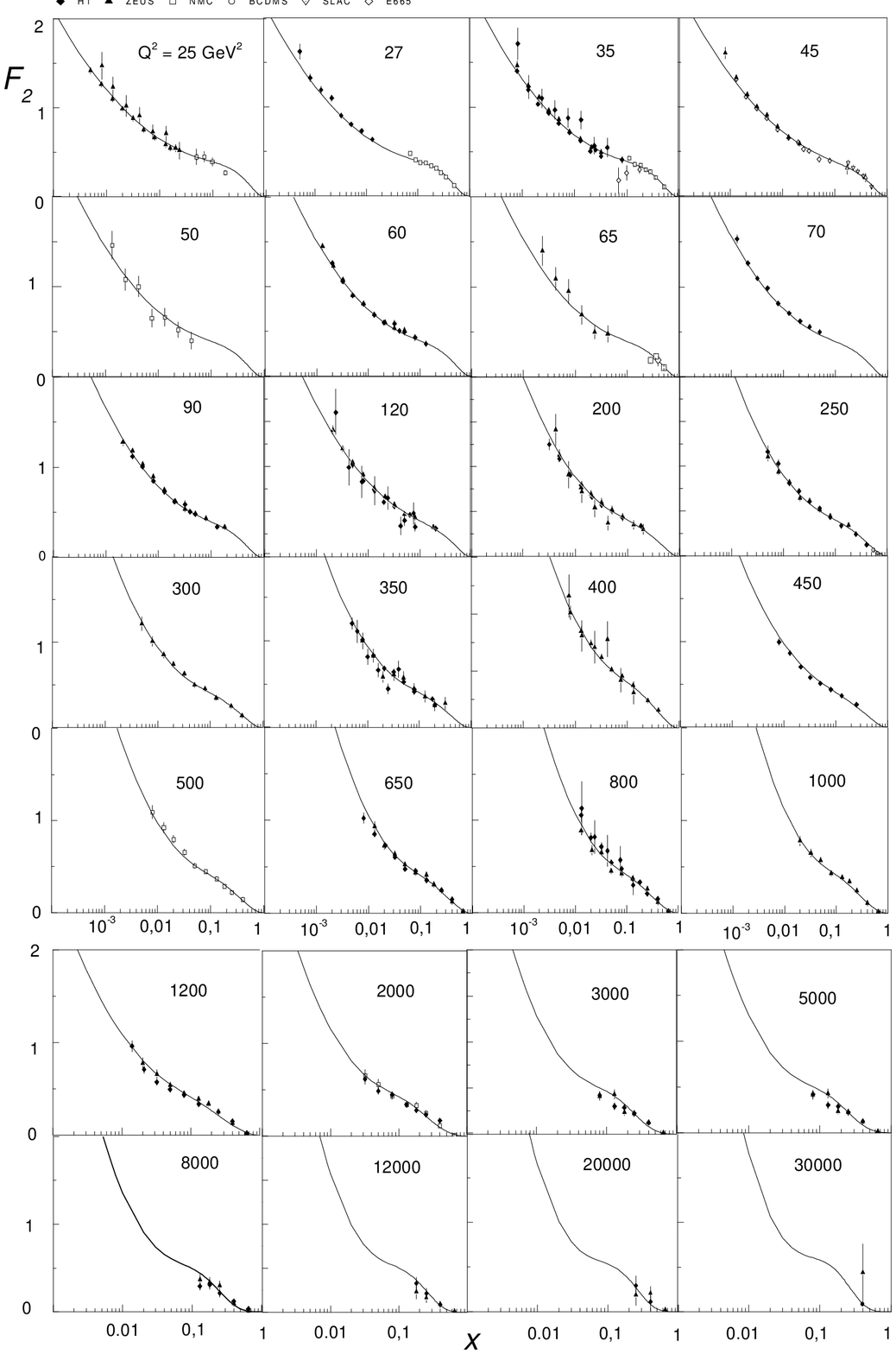,width=13cm,height=16cm}}
\caption{The results for  the inclusive structure function at
medium and high $Q^{2}$ virtualities (in units of GeV$^2$) } \label{fig2}
\end{figure}

\begin{figure}[h]
\centerline{\epsfig{file=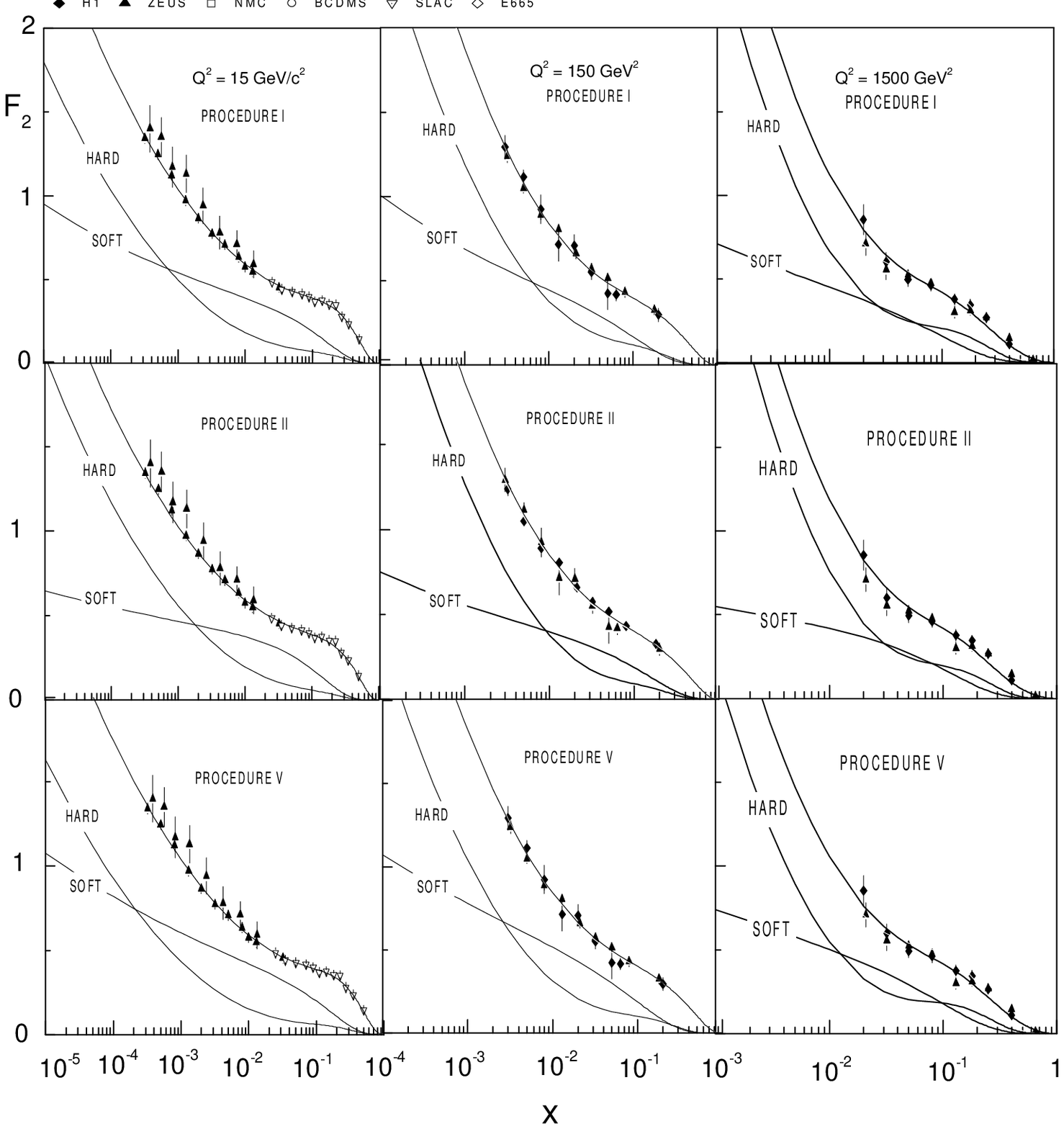,width=14cm,height=16cm}}
\caption{The results for the inclusive structure function at $Q^{2}$ = 15,
150 and 1500 GeV$^2$ virtualities with the contribution of soft and
hard pieces for the  different procedures.} \label{fig3}
\end{figure}

Additionally,  we performed the fit  restricting the  domain of
 the Bjorken variable to $x\leq0.07$. This procedure is similar to that considered in analysis of  Ref.\cite{DM}. For this purpose we restricted the number of free parameters
(equal to 10) for Soft+Hard Pomeron model,  as in the Donnachie-Landshoff model  reanalysis considered in  \cite{DM}. To this aim, we  fixed the
constant $\alpha_s=0.2$ as well as  the powers $n_s=7$ and $n_{ns}=3$. The following  procedures were considered:  (A): hard plus soft terms;  (B): hard + soft dipole pomeron terms; (C): hard + soft tripole
pomeron terms. Results of the fit are presented in Table 2. The quality of our
approach is similar   those ones found in Ref. \cite{DM}, at least in
this specific  domain, $x\leq0.07$. The parameters for the multipole Pomeron seem stable in this kinematical region. There is a change in the case (A), where the bare Pomeron intercept is higher than in the procedure (I) producing an usual soft result $\Delta \simeq 0.08$. The parameters for the hard piece remain stable in all fitting procedures.

 In conclusion, we verify that the fitting
        procedure is equivalent to the model using the two-Pomeron approach
         \cite{DM,DL}, with the advantage of  clear understanding of the
         behaviors on $x$ and $Q^2$ of the corresponding hard content. In particular, the behavior on $Q^2$ of the hard piece sheds light about on further implementations of the hard Pomeron residue in two-Pomeron fitting. The fitting procedure shows that the soft Pomeron has an intercept small than the usual $\alpha_{\pom}=1.08$, suggesting that a suitable choice is the multipole pomeron having intercept equal to one. In particular, the tripole pomeron presented the best fit result in all kinematical ranges considered here, followed by the dipole one. The fitting using only the hard piece and the non-singlet contributions is not completely ruled out, producing a not so high $\chi^{2}/\mathrm{dof}$ result.

\section*{Acknowledgments}
M.V.T.M. thanks the support of the  High Energy Physics Phenomenology
Group (GFPAE, IF-UFRGS) at Institute of Physics, Porto Alegre.


\begin{thebibliography}{99}


\bibitem{MKlein} M. Klein. {\it Int. J. Mod. Phys.} {\bf
A15S1}, 467 (2000).


\bibitem{dippom}  P. Desgrolard, A. Lengyel, E. Martynov, {\it Eur. Phys. J.}
{\bf C7}, 655 (1999).\\
 P. Desgrolard,  L. Jenkovszky, F. Paccanoni, {\sl Eur. Phys. J.} {\bf C7}, 263
(1999).

\bibitem{DM} P. Desgrolard, E. Martynov,  {\sl Eur. Phys. J.} {\bf C22}, 479
(2001).

\bibitem{Cudell} J.R. Cudell, G. Soyez, {\it Phys. Lett.} {\bf B516},
77 (2001).


\bibitem{DL} A. Donnachie, P.V. Landshoff,  {\it Phys. Lett.} {\bf B518},
63 (2001), and references therein.

\bibitem{csernai} L. Csernai, L. Jenkovszky, K. Kontros, A. Lengyel, V. Magas, F. Paccanoni, {\it Eur. Phys. J.} {\bf C24},  205 (2002).

\bibitem{soyez} G. Soyez, e-Print Archive: hep-ph/0211361.


\bibitem{DGLAP} Yu.L.
Dokshitzer. {\it Sov. Phys. JETP} {\bf 46}, 641  (1977); \\ G. Altarelli and
G. Parisi. {\it Nucl. Phys.} {\bf B126}, 298  (1977); \\ V.N. Gribov and L.N.
Lipatov. {\it Sov. J. Nucl. Phys} {\bf 28}, 822  (1978).

\bibitem{OPEbreak} A.H. Mueller. {\it Phys. Lett.} {\bf  B396}, 251 (1997).

\bibitem{LOBFKL}E.A. Kuraev, L.N. Lipatov and V.S. Fadin.
 {\it Phys. Lett} {\bf B60} 50
(1975); {\it idem}, {\it Sov. Phys. JETP} {\bf 44} 443 (1976);
{\it Sov. Phys. JETP} {\bf 45} 199 (1977);  Ya. Balitsky and L.N.
Lipatov. {\it Sov. J. Nucl. Phys. } {\bf 28} 822 (1978).

\bibitem{BFKLNLO} V.S. Fadin, L.N. Lipatov, {\it Phys. Lett.} {\bf B429}, 127 (1998); M. Ciafaloni, G. Camici, {\it Phys. Lett.} {\bf B430}, 349 (1998)

\bibitem{Salam} G.P. Salam. {\it Acta Phys. Pol.} {\bf B30}, 3679 (1999).

\bibitem{Trunkbfkl} M.B. Gay Ducati, M.V.T. Machado. {\it Phys. Rev.} {\bf D63}, 094018 (2001);  {\it Nucl. Phys.} (Proc. Suppl.) {\bf B99}, 265 (2001).\\
M.B. Gay Ducati, M.V.T. Machado,  e-Print Archive: hep-ph/0104192.

\bibitem{Fiore} R. Fiore {\it et al}., {\it Phys. Rev.} {\bf D63}, 056010
(2001).

\bibitem{PLB2002} M.B. Gay Ducati, K. Kontros, A. Lengyel, M.V.T. Machado, {\it Phys. Lett.} {\bf B533}, 43 (2002).

\bibitem{Predazzibook} V. Barone, E. Predazzi, {\it High Energy Particle Diffraction}, Springer-Verlag (2002).


\bibitem{LevRysk} E.M. Levin, M.G. Ryskin, {\it Sov. J. Nucl. Phys.} {\bf 53}, 653 (1991).

\bibitem{Balitsky} Ya. Balitsky, E. Kuchina, {\it Phys. Rev.} {\bf D62}, 074004
(2000).

\bibitem{KMP} A.B. Kaidalov, C. Merino, D. Pertermann, {\it Eur. Phys. J.} {\bf C 20}, 301 (2001).

\bibitem{CKMT} A. Capella, A.B. Kaidalov, C. Merino, J. Tran Thanh Van, {\it
Phys. Lett.} {\bf B337}, 358 (1994).

\bibitem{H1c3} H1 Collaboration, T. Ahmed {\sl et al.}, {\it Nucl. Phys.} {\bf B439}, 471
(1995).
\bibitem{H1c4} H1 Collaboration, S. Aid {\sl et al.}, {\it Nucl. Phys.} {\bf B470}, 3
(1996).
\bibitem{H1c5} H1 Collaboration, C. Adloff {\sl et al.}, {\it Nucl. Phys.} {\bf B497}, 3
(1997).
\bibitem{H1c6} H1 Collaboration, C. Adloff {\sl et al.}, {\it Eur. Phys. J.} {\bf C13}, 609
(2000).
\bibitem{H1c1} H1 Collaboration, C. Adloff {\it et al.},  {\sl Eur. Phys. J.}
{\bf C21}, 33 (2001).
\bibitem{H1c2} H1 Collaboration, C. Adloff {\it et al.}, {\sl Phys. Lett.}
{\bf B520}, 183 (2001).
\bibitem{ZEUSc1} ZEUS Collaboration, M. Derrick {\sl et al.}, {\it Zeit. Phys.} {\bf C72}, 399
(1996).
\bibitem{ZEUSc2} ZEUS Collaboration, J. Breitweg {\sl et al.}, {\it Phys. Lett.} {\bf B407},
432 (1997).
\bibitem{ZEUSc3} ZEUS Collaboration, J. Breitweg {\sl et al.}, {\it Eur. Phys. J.} {\bf C7},
609 (1999).
\bibitem{ZEUSc4} ZEUS Collaboration, J. Breitweg {\sl et al.}, {\it Nucl. Phys.} {\bf B487}, 53
(2000).
\bibitem{ZEUSc5} ZEUS Collaboration, S. Chekanov {\sl et al.}, {\it Eur. Phys. J.}
{\bf C21}, 443 (2001).
\bibitem{E665} E665 collaboration, M.R. Adams{\sl et al.},
{\sl Phys. Rev.}{\bf D54}, 3006 (1996).
\bibitem{NMC} NMC collaboration, M. Arneodo {\sl et al.} {\it Nucl. Phys.} {\bf B483}, 3
(1997).
\bibitem{SLAC} SLAC old experiments, L.W. Whitlow {\sl et al.} {\it Phys.
Lett.}{\bf B282}, 475 (1992).
\bibitem{BCDSM} BCDSM Collaboration, A Benvenuti {\sl et al.} {\it Phys. Lett.} {\bf B223}, 485 (1989).

\end{thebibliography}
\end{document}